\documentclass{article}
\usepackage{spconf,amsmath,graphicx}
\usepackage{multirow}
\usepackage{enumitem}
\setlist{nosep, leftmargin=14pt}

\usepackage{mwe} 
\usepackage{booktabs}
\usepackage{multirow}
\usepackage[colorlinks=true, allcolors=green]{hyperref}
\usepackage{amsmath, amssymb}

\title{Neural Radiance Fields in Medical Image: A Survey}
\name{
Xin Wang$^{1, \star}$, \thanks{ $\star$ Corresponding author.} 
Yineng Cheng$^{1}$,
Shu Hu$^{2}$,
Heng Fan$^{3}$,
Hongtu Zhu$^{4}$,
Xin Li$^{1}$
}
\address{
 $^{1}$University at Albany, State University of New York  
 {\tt\{xwang56,ychen77,xli48\}@albany.edu} \\
 $^{2}$Purdue University 
 {\tt hu968@purdue.edu}  \\
 $^{3}$University of North Texas 
 {\tt heng.fan@unt.edu }  \\
 $^{4}$University of North Carolina at Chapel Hill
 {\tt htzhu@email.unc.edu } 
}
\begin{document}
%
\maketitle

\thispagestyle{plain}
\pagestyle{plain}

\begin{abstract}
Neural Radiance Fields (NeRF), as a pioneering technique in computer vision, offer great potential to revolutionize medical imaging by synthesizing three-dimensional representations from projected two-dimensional image data. 
However, they face unique challenges when applied to medical imaging. 
This paper presents a comprehensive examination of NeRF applications in medical imaging, highlighting four imminent challenges, including fundamental imaging principles, internal structural requirement, object boundary definition, and color density significance. We discuss existing methods applied to different organs and their associated limitations. We also review several datasets and evaluation metrics and propose promising directions for future research.

\end{abstract}

%
%
%
%
\section{Introduction}
\label{sec:intro}


Neural Radiance Fields (NeRFs) significantly enhance traditional 3D reconstruction methods in several key aspects \cite{mildenhall2021nerf}. 
Unlike conventional techniques that often rely on geometric or voxel-based approaches, NeRFs utilize deep learning to create highly detailed and photorealistic 3D models from a set of 2D images \cite{zhu2023deep}. 
Compared with traditional reconstruction methods,  NeRFs can render complex scenes with intricate light and shadow interplay, which traditional methods struggle to capture accurately \cite{nguyen2025dwtnerf}. 
They also excel in handling occlusions and view-dependent effects, producing more continuous and smoother representations of 3D spaces. 
Additionally, NeRFs require fewer manual interventions and assumptions about scene geometry, leading to more automated and scalable 3D reconstructions \cite{rabby2023beyondpixels}. 
This capability is particularly advantageous in medical imaging, where precise and detailed visualization of anatomical structures is crucial, offering enhanced diagnostic and surgical planning tools over traditional methods \cite{lo2012extraction}.

\begin{figure}[t] %
\centering
\includegraphics[width=0.48\textwidth]{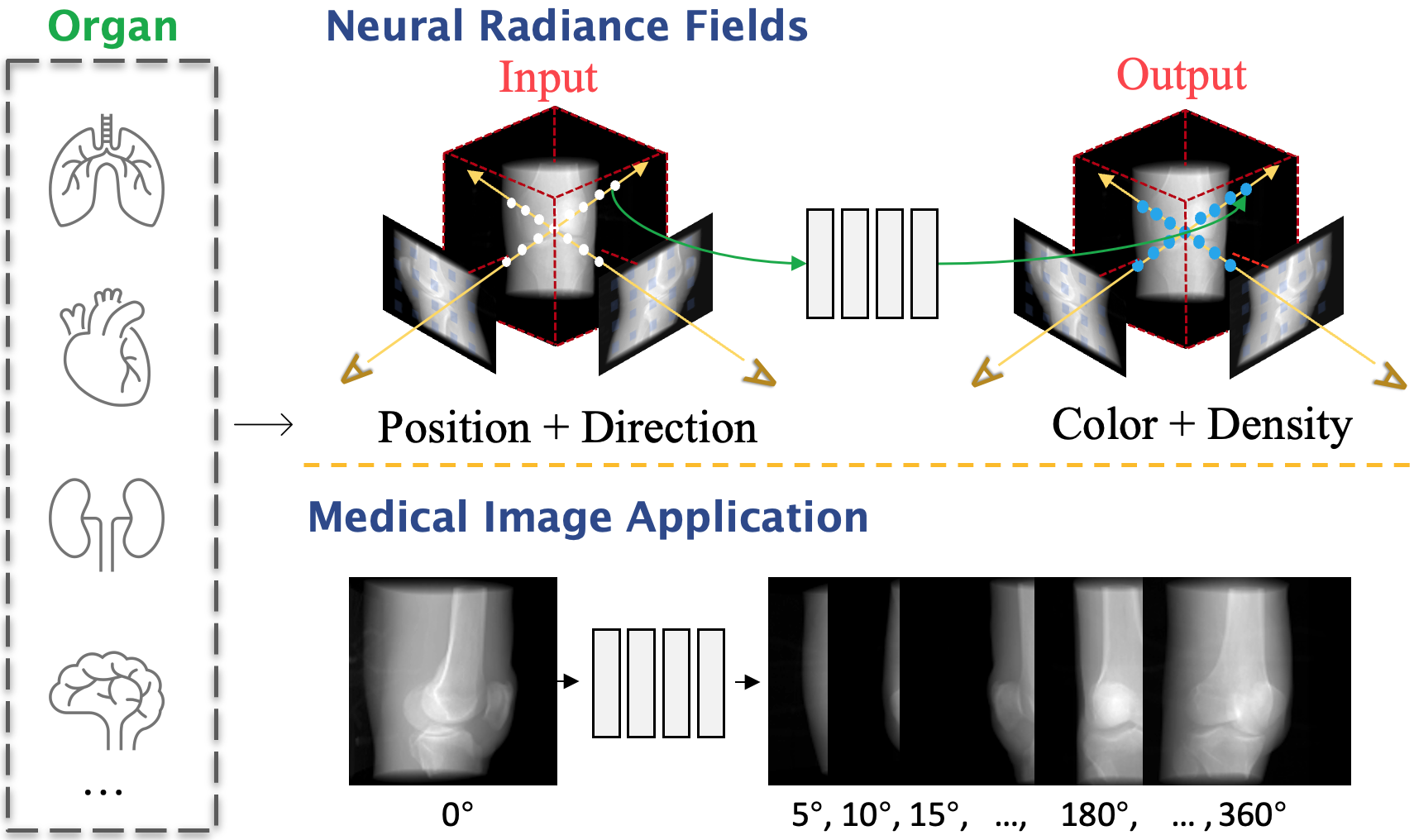}
\vspace{-5mm}
\caption{\small \it Overview of this work, discuss NeRF for medical imaging of different organs. \textbf{Top:} NeRF model \cite{mildenhall2021nerf}. 
\textbf{Bottom:} An application of NeRF in medical images, the goal is to render the 3D volume from a single view \cite{corona2022mednerf,hu2023umednerf}. 
}
\label{figintro}
\vspace{-5mm}
\end{figure}

%
By employing deep learning techniques to synthesize highly detailed and photorealistic 3D models from sets of 2D medical images, such as CT scans or MRIs, NeRFs provide an unprecedented level of detail and accuracy in visualizing complex anatomical structures \cite{mildenhall2021nerf,wysocki2024ultra}. 
This technique surpasses traditional 3D reconstruction methods by effectively handling intricate light interactions and view-dependent phenomena, thus providing clearer, more continuous representations of internal body structures \cite{song2023oral}. 
Furthermore, NeRFs can potentially minimize the need for multiple or high-resolution scans that expose patients to excessive ionizing radiation \cite{corona2022mednerf,dagli2024nerf}. 
Although the NeRFs-based method has yielded surprising results in traditional 3D image reconstruction, it cannot be directly applied to medical image reconstruction due to the unique internal details of medical images \cite{molaei2023implicit,rabby2023beyondpixels}.


This paper aims to highlight the significant motivation and challenges associated with implementing NeRFs in medical imaging. 
We have systematically categorized existing methods based on organ types, acknowledging that the distinct characteristics of various organ structures, such as sparsity, overlap, and visibility, pose considerable hurdles to achieving comprehensive morphological reconstructions. 
These obstacles often lead to incomplete 3D models, potentially omitting crucial details (see Figure \ref{figintro} for an example). 
Then, we explore the future research opportunities that could help overcome these obstacles and advancing the field.

The key contributions of this work are as follows:

\begin{itemize}
\item We outline the key motivations and challenges associated with applying NeRFs to medical imaging, identifying the main hurdles that must be overcome.

\item We categorize and review the existing NeRF methodologies based on specific organs, highlighting their limitations and the unique challenges of different anatomical structures.

\item Furthermore, we explore potential future research avenues, emphasizing how NeRFs in medical imaging can revolutionize diagnostic processes, surgical planning, and medical education. This represents a significant advancement in medical visualization and patient care, promising to enhance the accuracy and efficacy of treatments.
\end{itemize}



The remaining content of this paper is organized as follows. After introducing the motivation and challenges of NeRFs in medical imaging in Section~\ref{Challenges}, 
we describe the medical image NeRF applications for different organs and discuss the limitations of the current methods in Section~\ref{method}. 
In Sections \ref{Dataset}, we summarize the public datasets for NeRFs in medical imaging. 
We discuss the future directions in Section \ref{futuredirection} and conclude the paper in Section~\ref{conclusion}.

\section{Motivation and Challenges}
\label{Challenges}

\subsection{Motivation}





Neural Radiance Fields (NeRFs) in medical imaging present innovative objectives to enhance imaging techniques in healthcare, addressing several key clinical needs.

\vspace{0.3em}
\noindent
\textbf{(1) Minimizing Radiation Exposure.} 
NeRFs in medical imaging can potentially reduce the necessity for multiple or high-resolution scans that subject patients to high levels of ionizing radiation. By generating 3D representations from a sparse collection of 2D images, these methods can decrease the frequency of exposures, lowering radiation doses and reducing the risks associated with radiation exposure. This technique not only reduces the risk of radiation-related health conditions but it also alleviates issues related to prolonged patient immobility during imaging sessions.

\vspace{0.3em}
\noindent
\textbf{(2) Reducing Time Requirements.} 
Compiling multiple scans into a coherent 3D representation is time-intensive, posing significant challenges in time-sensitive medical situations. NeRFs in medical imaging streamline this process by enabling the quick creation of 3D models from a limited number of images, thereby accelerating data collection and image reconstruction. This rapid processing is particularly advantageous in emergency medicine settings, where efficiency can directly impact patient care and outcomes.

\vspace{0.3em}
\noindent
\textbf{(3) Lowering Imaging Costs.} 
Achieving detailed 3D images with conventional CT scans requires multiple X-ray projections, each with a fine slice thickness, which can be costly. NeRFs in medical imaging address this issue by efficiently producing detailed 3D models from smaller images, potentially reducing the financial burden. 
The capability of NeRFs in medical imaging to deliver precise volumetric reconstructions with fewer data inputs suggests a decrease in the reliance on expensive imaging resources, offering a cost-effective alternative to traditional imaging methods.

\vspace{0.3em}
\noindent
\textbf{(4) Reducing Motion-related Artifacts.} Medical imaging reconstructions typically involve multiple scan acquisitions, causing lengthy sessions that necessitate patient stillness, thereby increasing the risk of motion artifacts. Specifically, Magnetic Resonance Imaging (MRI) sequences are highly susceptible to motion-related artifacts caused by breathing and cardiac pulsation, potentially leading to inaccuracies in 3D reconstruction. Such inaccuracies are problematic in clinical settings, where precise anatomical measurements are essential for accurate treatment planning and patient monitoring.


\begin{figure}[t] %
\centering
\includegraphics[width=0.45\textwidth]{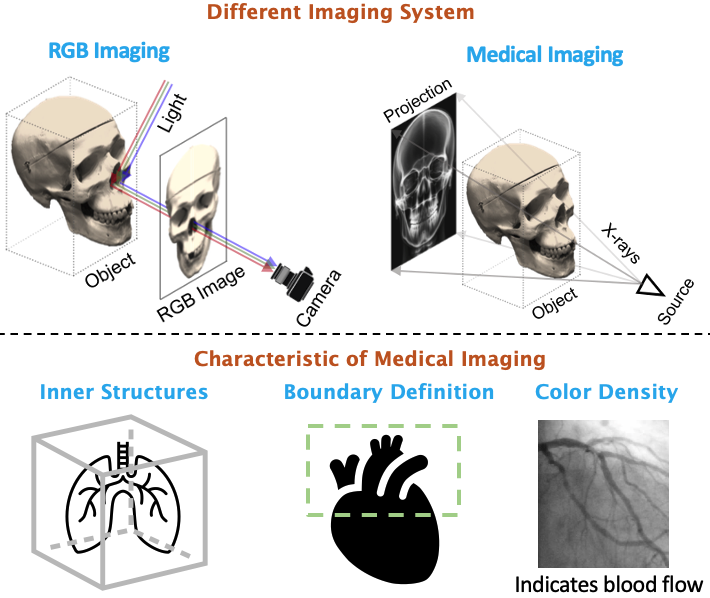}
\vspace{-1mm}
\caption{\small \it Overview of the challenges of NeRFs in medical images. \textbf{Top:} The fundamental imaging principles of RBG image and medical image are different, an example of X-rays imaging is from \cite{cai2023structure}. \textbf{Bottom:} (left) Medical images necessitate detailed inner structure visualization. (mid) often suffer from poor object boundary definition, and (right) exhibit the varying significance of color density. }
\label{figchallenge}
\vspace{-3mm}
\end{figure}

\subsection{Challenges}

Despite promising performance in reconstructing 3D natural images, Neural Radiance Fields (NeRFs) face several challenges when applied to medical imaging, primarily due to the unique characteristics and requirements of medical image data. An overview of these challenges is shown in Figure \ref{figchallenge}.

\vspace{0.3em}
\noindent
\textbf{(1) Different Imaging Principles.} 
The fundamental imaging principles of consumer electronics and medical imaging are different. For example, while color-based (RGB) NeRF methods have seen considerable development, their direct application to X-ray imaging may yield suboptimal outcomes due to inherent differences between visible light and X-ray modalities. Unlike visible light imaging, which captures external features by reflecting off an object's surface, X-ray imaging penetrates and attenuates the object, creating images that primarily reveal internal structures. These internal details are crucial for 3D X-ray reconstruction, highlighting the need for tailored NeRF algorithms to effectively accommodate the unique characteristics of X-ray imaging \cite{cai2023structure}.

\vspace{0.3em}
\noindent
\textbf{(2) Need for Detailed Inner Structures.} Medical images often require high levels of detail to represent intricate inner structures of the body accurately. To be clinically useful, NeRFs, which synthesize 3D models from 2D images, must capture these details. However, accurately rendering the complex internal anatomy, such as the fine distinctions between different types of tissues or the subtle anomalies indicative of early-stage diseases, can be challenging. This is due to the limitations in the resolution and the interpolation capabilities of current NeRF models, which may struggle to replicate the exact internal textures and structures needed for precise medical diagnosis and treatment planning.

\vspace{0.3em}
\noindent
\textbf{(3) Ambiguity of Object Boundary.} Medical images frequently run into ambiguity in object boundaries between different types of tissues or between pathological and healthy tissues. This ambiguity in boundary delineation poses a significant challenge for NeRFs, as these models rely on clear distinctions in visual data to reconstruct 3D spaces accurately. In medical imaging, where precise boundary identification is crucial for tasks such as tumor localization or the measurement of organ volumes, NeRFs must overcome this hurdle to ensure the reliability and accuracy of generated 3D models.


\noindent
\textbf{(4) Significance of Color Density.} In medical imaging, variations in color density can convey critical information, such as the presence of blood flow (see Figure \ref{figchallenge}), variations in tissue density, or the accumulation of contrast agents. 
These variations are essential for diagnosing conditions and planning treatments. NeRFs, which inherently model the light transport in a scene, must be adapted to interpret and accurately reproduce these color density variations in the synthesized 3D images. The challenge lies in the fact that NeRFs were initially designed for visual scenes where lighting plays a significant role in the appearance of objects, rather than for medical scenarios where color density variations are more closely related to organ properties and physiological conditions.

\section{Methods}
\label{method}



The unique properties of various organ structures introduce significant challenges to the reconstruction process, such as sparsity, overlap, and visibility issues \cite{yang2024efficient}, which can hinder the creation of complete morphological models, potentially resulting in missing details in the 3D reconstructions \cite{liu2025neural}. 
A prime example is the vessel structure,  especially the challenge of vessel sparsity, where the thin nature of vessels requires a thorough sampling methodology for accurate depiction. 
Consequently, this section provides a comprehensive classification based on the application of NeRFs across different organs. 
Our goal is to acquaint researchers with the essential operations and functionalities of NeRF models in medical imaging, addressing the intricacies involved in capturing the full complexity of organ structures.

\begin{figure}[t] %
\centering
\includegraphics[width=0.48\textwidth]{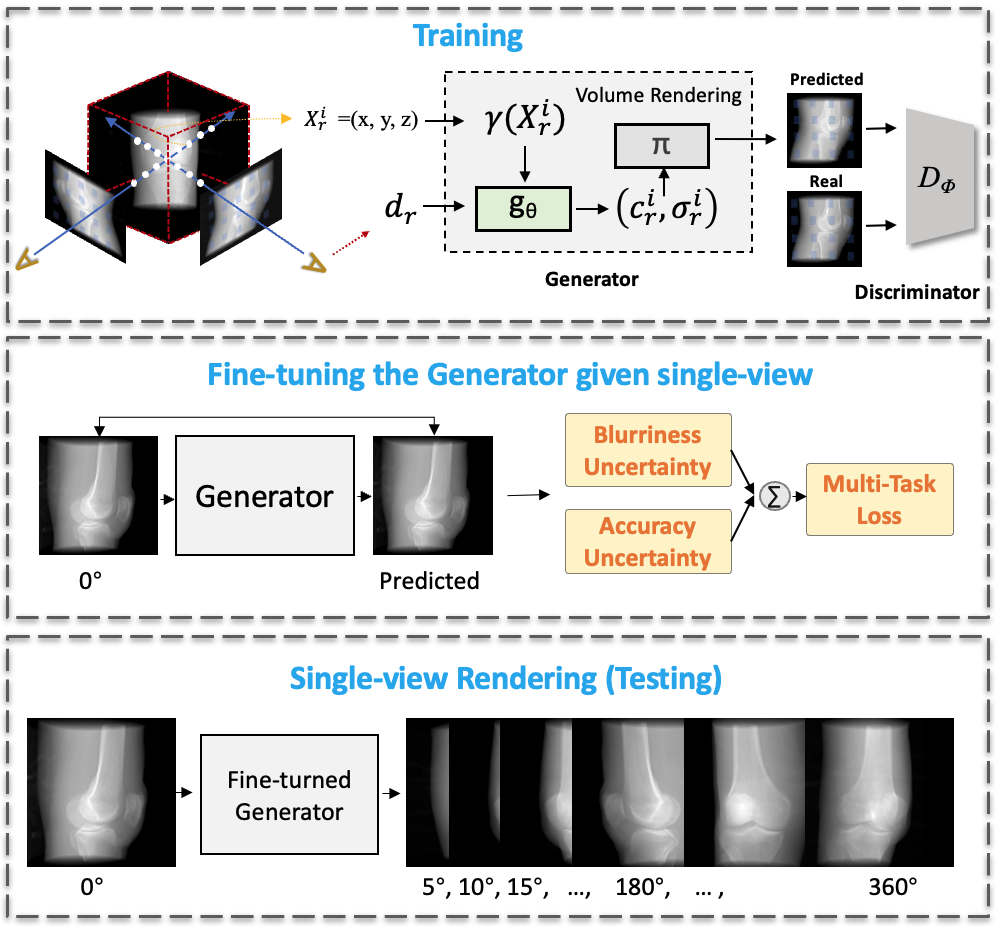}
\vspace{-6mm}
\caption{\small \it Over view of Uncertainty-aware MedNeRF \cite{hu2023umednerf}. 
\textbf{Top:} Network structures and training phrase \cite{corona2022mednerf}. 
\textbf{Mid:}  Fine-tuning the generator for balancing the blurriness and accuracy with uncertainty-aware multi-task loss. 
\textbf{Bottom:} Rendering the whole 3D volume (each view taken at a 5-degree interval) using a single view slide.}
\label{umednerf}
\vspace{-3mm}
\end{figure}

\subsection{NeRFs for Different Organs}


\noindent
\textbf{Knee.} 
Knee injuries and degenerative conditions, such as osteoarthritis, are very common \cite{bien2018deep}. Improved imaging and 3D reconstruction techniques can greatly aid in the diagnosis, treatment planning, and monitoring of such conditions, making the knee a highly relevant subject for research and application of NeRF technology.
%
Recently, MedNeRF \cite{corona2022mednerf}, a GAN-based Radiational Filed, successfully untangles the surface shape, volumetric depth, and internal anatomical structures from 2D images by learning a function that assigns a radiance value to each pixel. 
However, due to the nature of GAN-based network architecture and  the reliance on a single X-ray for CT projection reconstruction, the network may overfit to the single X-ray, resulting in an imbalance between image clarity and accuracy. 
Thus, during the single-view volumetric rendering stage, MedNeRF fine-tuned the generator using the input view by jointly optimizing the reconstruction and the generator. 
The model showcases high-fidelity renderings, both qualitatively and quantitatively, and is compared favorably against other recent methods based on radiance fields, highlighting its effectiveness in reducing radiation exposure while maintaining imaging quality \cite{sun2024acnerf}.
However, the hyperparameters in MedNeRF for balancing two losses (clarity and accuracy) are typically configured manually and subject to observation, and heavily depend on human expertise, which can pose challenges in achieving the optimal trade-off between image blurriness and accuracy. 
To address this challenge, Uncertainty-aware MedNeRF ({UMedNeRF}) \cite{hu2023umednerf} is presented to automatically learn hyperparameters for clarity and precision balancing problems. 
As shown in Figure \ref{umednerf}, 
the key to improving the quality of the generated images lies in simultaneously ensuring image resolution and accuracy. 
In that case, it becomes a multi-task learning problem, where an uncertainty-aware conditional radiance field that simultaneously weighs multiple loss functions automatically in the multi-task settings and generates the predicted patch for single-view volumetric rendering.


\begin{figure}[t] %
\centering
\includegraphics[width=0.45\textwidth]{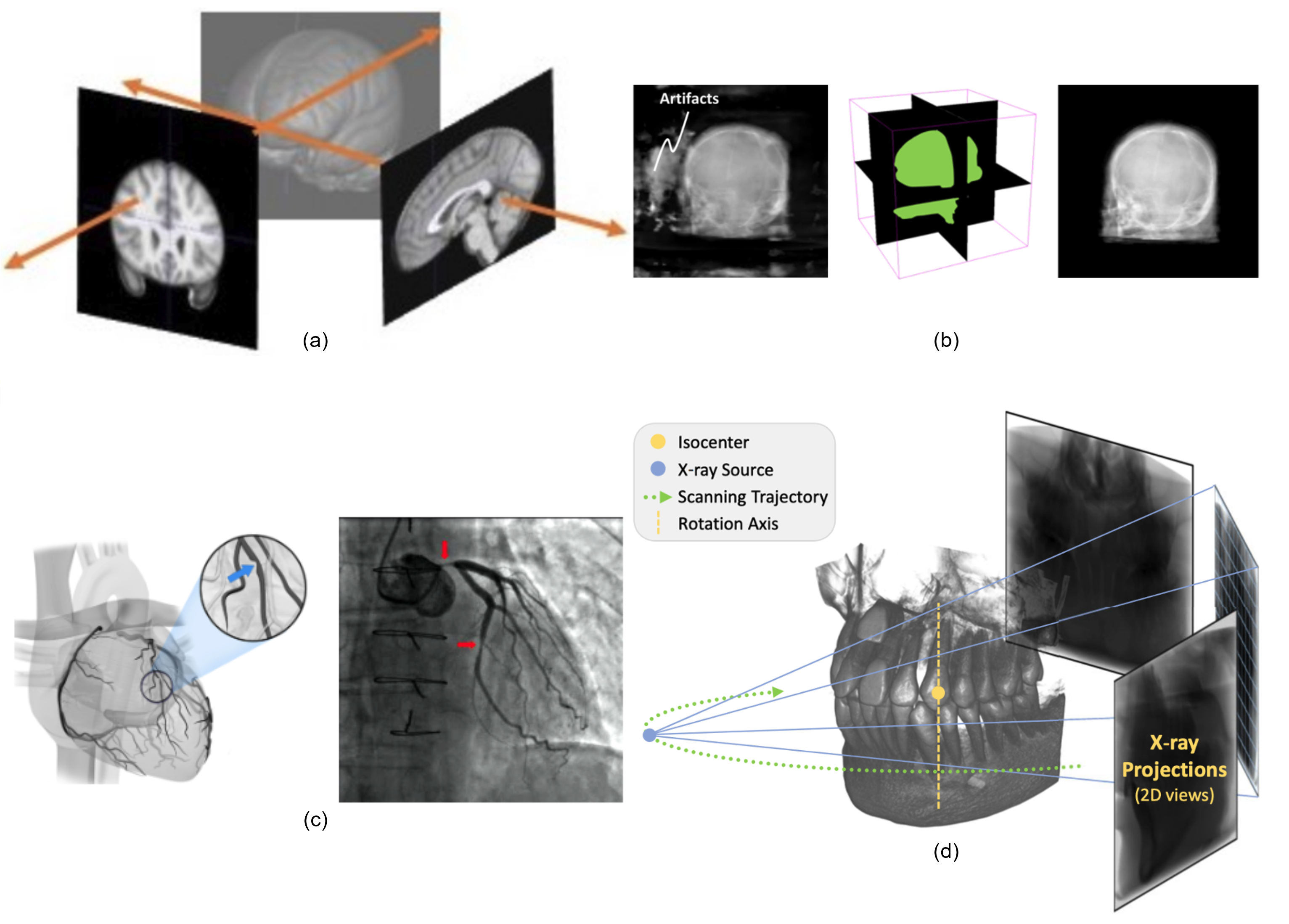}
\vspace{-1mm}
\caption{\small \it 
(a) Overview of extended-MedNeRF with 2D Slices of Brain MRI scans in \cite{iddrisu20233d}.
(b) Overview of Masked NeRF \cite{zhou2023robust}. \textbf{Left:} Skull projections from NeRF. \textbf{Right.} Skull projections from Masked NeRF scene representations.
(c)  Overview of Coronary Angiography   \cite{WikiCoronaryCath2023}.
(d) Overview of Sparse-view Neural Attenuation Fields (SNAF) \cite{fang2022snaf}. An X-ray source systematically traverses around the region of interest, producing sparse projections captured by a flat-panel detector positioned on the opposite side.
}
\label{fig_example}
\vspace{-3mm}
\end{figure}

\smallskip
\noindent
\textbf{Brain.} 
Brain Magnetic Resonance Imaging (MRI) is pivotal in medical diagnostics and research because it allows for the detailed visualization of brain structures, aiding in the diagnosis, monitoring, and treatment planning of neurological conditions. 
Utilizing advanced imaging techniques, e.g., NeRFs, can enhance the quality and accuracy of brain images, and thus improve the detection of abnormalities, such as tumors, lesions, and degenerative diseases, providing crucial information for clinical decision-making and patient care. 
NeRFs can contribute to the development of more precise and less invasive diagnostic methods, improving patient outcomes and advancing the understanding of brain anatomy and function.
Specifically, the advent of NeRFs offers a promising solution, enabling the generation of precise 3D reconstructions from MRI images with minimal user input. Leveraging 3D convolutional neural networks, a recent study \cite{iddrisu20233d} reconstructs 3D projections from 2D MRI slices, overcoming challenges such as variable slice thickness and integrating multiple MRI modalities for enhanced robustness (see Figure \ref{fig_example} a). 
This technique facilitates accurate slice interpolation, capturing the shape and volumetric depth of both surface and internal structures.  It also improves the accuracy and reliability of 3D anatomical reconstructions, offering medical professionals the ability to visualize and analyze complex structures with fewer scans, potentially decreasing imaging times and mitigating motion-related complications.


\smallskip
\noindent
\textbf{Skulls.} 
Skull Cone Beam Computed Tomography (CBCT) scans are the gold standard for diagnosing cranial fractures. They provide high-resolution images of the bone, allowing clinicians to identify even minor fractures that could be critical to a patient's treatment plan.
In the realm of image-guided, minimally invasive medical procedures, accurate pose estimation plays a pivotal role, utilizing X-ray projections to precisely navigate to targets within 3D space. 
%
An unconstrained NeRF model applied to skull CBCT reconstructions exhibited artifacts beyond the reconstructed anatomical boundaries \cite{zhou2023robust}. 
These artifacts stem from overfitting to a constrained set of ground-truth X-ray images, resulting in implausible visualizations for novel poses. 
To mitigate these issues, a Masked-NeRF (mNeRF) approach is introduced by leveraging a 3D head mask derived from CBCT scans to spatially constrain NeRF's extrapolations. The mask, deliberately expanded by 3mm for comprehensive coverage, ensures that densities beyond its perimeter are negated, facilitating seamless integration within the deep learning framework for gradient-based optimization and processing (see Figure \ref{fig_example} b).


\smallskip
\noindent
\textbf{Heart Vessel.} 
Cardiac vessel imaging is crucial for diagnosing various heart conditions, including coronary artery disease (CAD), aneurysms, and blockages \cite{wang2024artificial,kshirsagar2024generative}. 
It allows doctors to assess the health of the heart's blood vessels and identify areas of reduced blood flow. An example of coronary angiography is shown in Figure \ref{fig_example} c.
However, the inherent characteristics of vessel structures present additional challenges, such as sparsity, overlap, and visibility, which obstruct the comprehensive reconstruction of vascular morphologies, potentially leading to omissions in the 3D models. These challenges include (1) vessel sparsity, which refers to the slender nature of vessels necessitating a comprehensive sampling approach for accurate reconstruction, (2) vessel overlap, concerning the concealment of crucial vessel segments in 2D projections due to occlusions, and (3) vessel visibility issues, where parts of the vessel may be obscured due to contrast variability or the superimposition of background structures. 
A recent study \cite{maas2023nerf} is dedicated to addressing the challenges inherent in the 3D reconstruction of coronary angiography images and exploring the applicability of the NeRF in overcoming these obstacles. 
Existing methods for the 3D reconstruction of coronary arteries often fall short, hampered by issues such as vessel morphology characteristics including sparsity, overlap, and distinguishing between foreground and background, alongside constraints posed by sparse-view and limited-angle X-ray projections. 
The results indicate NeRF's promising capabilities for 3D reconstruction in the context of sparse and limited angle X-ray projections, while also pinpointing specific limitations, such as background overlap, that necessitate further research. 
Note that the method for large vessel Aorta is introduced in \cite{zha2022naf}.



\smallskip
\noindent
\textbf{Teeth.} Dental Cone Beam Computed Tomography (CBCT) images enable the detection and diagnosis of various dental conditions such as cavities, tooth decay, and gum disease \cite{liu2024geometry,zou2024pa,li20243dpx}. 
They provide detailed views that are not visible to the naked eye, allowing for early treatment.
Neural Attenuation Fields (NAF) \cite{zha2022naf} innovatively model the attenuation coefficients as a continuous function across 3D space, utilizing a dense neural network for parameterization. 
A novel learning-based encoder, incorporating hash coding, enhances the model's ability to delineate high-frequency details more effectively than traditional frequency-domain encoders. 
Furthermore, efforts to reconstruct high-quality CBCT images from sparse-view 2D projections have been ongoing, yet current leading methods often yield images marred by artifacts and missing fine details (see Figure \ref{fig_example} d). 
Sparse-view Neural Attenuation Fields (SNAF) \cite{fang2022snaf} has been proposed for sparse-view CBCT reconstruction. The view augmentation strategy addresses the data insufficiency challenges posed by sparse input views. 

\smallskip
\noindent
\textbf{Foot.}  
Foot CT images are crucial for diagnosing a wide range of conditions, from fractures and dislocations to degenerative diseases like arthritis and conditions such as plantar fasciitis. Developing imaging techniques that can offer clear, detailed views of the foot's internal structures can significantly improve diagnostic accuracy and treatment planning \cite{zha2022naf}. 

\smallskip
\noindent
\textbf{Abdomen.}  Abdominal CT imaging is crucial for diagnosing diseases such as cancers, inflammatory conditions, and vascular disorders. It helps in detecting tumors, assessing organ damage, identifying blockages or abnormalities in the intestines, and evaluating the extent of diseases (staging), which is essential for treatment planning \cite{zha2022naf}.

\smallskip
\noindent
\textbf{Chest.} Chest CT imaging is a cornerstone of medical diagnostics due to its ability to non-invasively reveal critical information about major organs, its role in a wide array of clinical applications, and its continual evolution alongside imaging technology advancements.
The chest contains critical organs such as the lungs, heart, and major blood vessels. 
Imaging these structures can help diagnose a wide range of conditions, from respiratory infections and lung cancer to heart disease and vascular disorders. 
Chest images provide valuable insights into the anatomy and potential abnormalities within the thoracic cavity. 
The works in the previous section \cite{corona2022mednerf,hu2023umednerf,sun2024acnerf, zha2022naf} are also evaluated using the chest CT images.











\subsection{Limitations}
\label{Limitations}



Although impressive progress has been made in the performance of NeRFs in medical imaging, several concerns remain regarding the current methods.


\vspace{0.3em}
\noindent
\textbf{Dataset Quality and Diversity.}
NeRFs require a substantial amount of paired data for training, specifically diverse and high-quality images from various angles and perspectives. See Section \ref{Dataset} for more details. 
In medical imaging, obtaining such paired datasets can be difficult due to privacy concerns, the rarity of certain conditions and the technical and ethical challenges associated with collecting and sharing patient data.


\vspace{0.3em}
\noindent
\textbf{Limited Resolution and Detail.}
While NeRFs can generate impressive 3D reconstructions, capturing the intricate details necessary for medical diagnosisremains a challenge. 
The resolution and fidelity of NeRF-generated images may not yet match the standards required to visualize fine anatomical structures or pathological changes crucial for accurate medical assessments \cite{chen2023cunerf}.

\vspace{0.3em}
\noindent
\textbf{Handling of Transparent and Reflective Surfaces.}
Medical images often contain structures that are semi-transparent or exhibit varying degrees of reflectivity, such as soft tissues compared to bones. 
NeRFs, originally designed for opaque objects in natural light settings, may struggle to accurately model these properties, leading to less accurate reconstructions.

\vspace{0.3em}
\noindent
\textbf{Boundary Ambiguity and Artifacts.}
The accurate delineation of boundaries between different tissues or pathological versus healthy areas is critical in medical imaging. 
NeRFs may produce images with ambiguous boundaries or artifacts, especially when dealing with complex or overlapping structures, potentially leading to misinterpretations.

\vspace{0.3em}
\noindent
\textbf{Generalization Across Different Imaging Modalities.}
Medical imaging encompasses a wide range of modalities (e.g., MRI, CT, and X-ray), each with its own unique characteristics. NeRFs developed for one type of imaging may not generalize well to another without significant adjustments or retraining, limiting their versatility \cite{wysocki2024ultra}.

\vspace{0.3em}
\noindent
\textbf{Interpretability and Clinical Validation.}
For NeRFs to be adopted in clinical practice, their outputs must be interpretable to medical professionals and validated against established medical imaging standards \cite{wang2022neural}. 
The black-box nature of deep learning models, including NeRFs,  make it challenging to understand how they generate their outputs, complicating clinical validation and trust.

\section{Datasets and Evaluation Metrics}
\label{Dataset}



\begin{figure}[t] %
\centering
\includegraphics[width=0.45\textwidth]{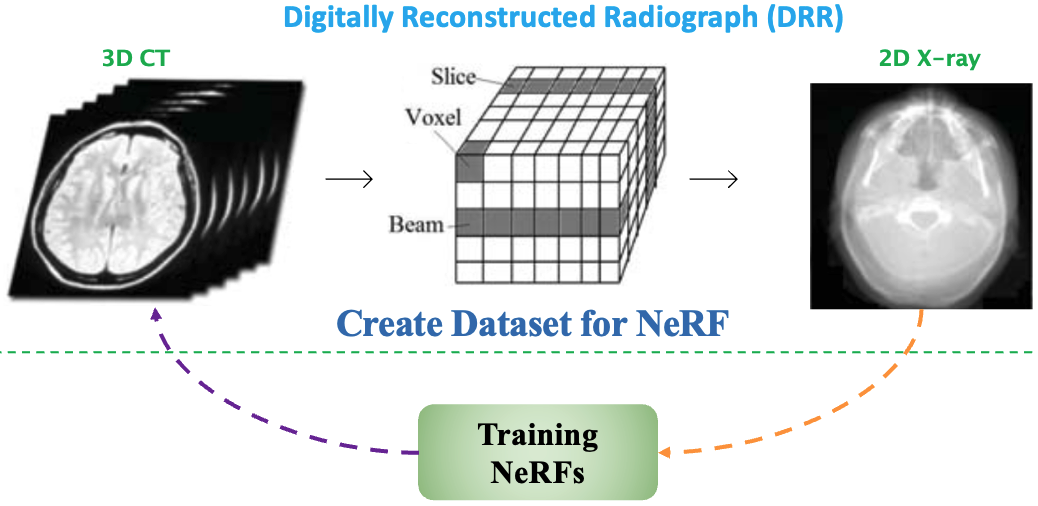}
\vspace{-1mm}
\caption{\small \it The relation between Digitally Reconstructed Radiograph (DRR) and medical imaging NeRFs, the DRR example is from  \cite{montufar2018perspective}. \textbf{Left:} leverages original CT scan slices, \textbf{Mid:} perceived as a three-dimensional voxel array, \textbf{Right:} to generate a DRR. 
This direct volume rendering scheme efficiently transforms the volumetric data from a CT scan into a visually comprehensive DRR, offering a synthesized view similar to traditional X-ray imagery. 
}
\label{figdrr}
\vspace{-5mm}
\end{figure}

Training Medical Image NeRFs needs datasets comprising paired 2D (e.g. X-ray) and 3D (e.g. CT, MRI) scans. However, assembling a real dataset of such pairs is virtually impractical due to the rarity of obtaining X-ray pairs and CT (or MRI) scans captured simultaneously in clinical practice \cite{10096296}. 
For example, gathering paired X-ray and CT data can introduce inaccuracies due to patient movement and variations in equipment. Additionally, this process may expose patients to increased radiation levels \cite{sun2024acnerf}.
Thus, the Reconstructed Radiographs (DRR) technique is employed for simulating and generating X-ray images \cite{corona2022mednerf}. 
This method entails the creation of virtual X-ray projections using three-dimensional medical image data, such as CT or MRI. 
By generating DRRs, we can avoid the need for paired X-rays and corresponding CT reconstructions, which would otherwise subject patients to increased radiation exposure.  
The overview relation between Digitally Reconstructed Radiographs (DRR) and medical imaging NeRFs is shown in Figure \ref{figdrr}.


\subsection{Digitally Reconstructed Radiographs (DRR)}


Digitally Reconstructed Radiographs (DRRs) are synthetic X-ray images generated from three-dimensional (3D) computed tomography (CT) data using digital image processing techniques. 
This technology allows for the creation of X-ray-like images without the need for additional radiation exposure to the patient. DRRs are produced by simulating the path of X-rays through a 3D volume dataset, mimicking the attenuation and absorption properties of X-rays as they pass through different tissues and structures within the body.
The primary utility of DRRs lies in their application within various medical fields, including radiation oncology, orthopedics, and interventional radiology, among others. In radiation therapy planning, for instance, DRRs are used to accurately target treatment areas, allowing for precise alignment of the therapeutic radiation beam with the tumor while sparing surrounding healthy tissue. 
DRR technology facilitates the comparison and verification of patient positioning during treatment, ensuring that the actual patient alignment matches the planned treatment setup. By leveraging existing CT datasets to generate these radiograph-like images, DRRs eliminate the need for additional diagnostic X-ray examinations, thereby reducing the cumulative radiation dose to the patient and enhancing the efficiency of medical imaging workflows. More DRR generation examples can be found in \cite{montufar2018perspective}.


\subsection{Public Dataset}

\noindent
\textbf{MedNeRF Dataset \cite{corona2022mednerf} (Github \cite{mednerfdataset}).} 
The dataset includes 20 CT chest scans as  referenced in \cite{tsai2021medical} and \cite{clark2013cancer}, in addition to 5 CT knee scans documented in \cite{harris2016combined} and \cite{ali2016validation}. 
The DRR generation process involves the emulation of the rotation of a radiation source and imaging panel around the vertical axis, resulting in the creation of DRRs with a resolution of 128 $\times$ 128 at five-degree intervals. 
Consequently, each object produces a total of 72 distinct DRRs. 

\vspace{0.3em}
\noindent
\textbf{NAF Dataset \cite{zha2022naf} (Github \cite{zha_naf_cbct}).} This publicly available Cone Beam Computed Tomography (CBCT) dataset of human organs, covers the chest, jaw, foot, and abdomen. 
The chest dataset is sourced from the LIDC-IDRI collection, while datasets for the jaw, foot, and abdomen are obtained from the Open Scientific Visualization Datasets. 
Given that these datasets solely provide volumetric CT scans, the projections are produced using the tomographic toolbox TIGRE. 

\vspace{0.3em}
\noindent
\textbf{PerX2CT Dataset \cite{10096296} (Github \cite{perx2ct_2024}).}  
The dataset is based on the LIDC-IDRI dataset and utilizes DRR technology to generate synthetic X-rays from actual CT volumes, simulating the required paired data for studies. The dataset is constructed following the methodology in\cite{ying2019x2ct}. 


\vspace{0.3em}
\noindent
\textbf{DIF-Net Dataset \cite{lin2023learning} (Github \cite{dif_net_dataset}).}  
The dataset comprises a knee CBCT collection with 614 CT scans, distributed as follows: 464 for training, 50 for validation, and 100 for testing purposes. 
It also includes 2D projections created using DRR technology, each with a resolution of 256 $\times$ 256.


\subsection{Evaluation Metrics}

Various methods have been developed to evaluate the effectiveness of NeRFs in medical imaging applications.

\noindent\textbf{2D Rendering.}
Another evaluation of the model focuses on the task of 2D rendering, utilizing the Frechet Inception Distance (FID) and Kernel Inception Distance (KID) metrics for a comparison of image quality, where lower scores signify superior quality \cite{corona2022mednerf}. 
For example, the models \cite{corona2022mednerf,hu2023umednerf} demonstrate superior performance, outperforming baseline models, such as pixelNeRF \cite{yu2021pixelnerf} and GIRAFFE \cite{schwarz2020graf, niemeyer2021giraffe}, in terms of both FID and KID metrics across all tested datasets.

\vspace{0.3em}
\noindent\textbf{3D Projection Reconstruction.}
To assess the efficacy of the model in reconstructing CT projections, two principal image quality metrics: Peak Signal-to-Noise Ratio (PSNR) and Structural Similarity Index (SSIM) are used \cite{corona2022mednerf}. 
PSNR focuses on the accuracy of luminance, quantifying the variance between the original and the model-generated images. 
Meanwhile, SSIM offers a holistic evaluation, taking into account factors such as luminance, contrast, and structural integrity. 
Utilizing these metrics is essential for a precise evaluation of the model's performance in CT projection reconstruction, providing a reliable basis for the research outcomes. These metrics are widely recognized for their effectiveness in measuring the quality of image reconstructions and CT projections, underscoring their importance in the analysis.








\section{FUTURE DIRECTIONS}
\label{futuredirection}

Besides continuing to address the aforementioned limitations, we also envision several important directions for NeRFs in medical imaging that will receive more attention in the future.




\vspace{0.3em}
\noindent
\textbf{Enhanced Resolution.}
Future developments should focus on improving the resolution and detail that NeRFs can capture, especially for internal structures within the medical images. 
Advanced algorithms could be designed to more effectively learn fine-grained details from limited or sparse imaging data, potentially through the integration of higher-resolution imaging techniques or more sophisticated methods.

\vspace{0.3em}
\noindent
\textbf{Improved Boundary Delineation.}
Addressing the issue of poor boundary definition in medical images will be crucial. Research could explore the use of hybrid models that combine NeRFs with segmentation networks or the incorporation of edge-detection mechanisms within the NeRF architecture to better distinguish between different types of tissues and pathological versus healthy areas.

\vspace{0.3em}
\noindent
\textbf{Accurate Color and Density Representation.}
Since color and density variations convey vital information in medical images, future NeRF models could benefit from enhancements in accurately modeling and reproducing these variations. 
This could involve adapting NeRFs to more precisely interpret and generate images based on the organ properties and physiological conditions that these color densities represent, perhaps by integrating additional data sources or modalities that provide complementary information.

\vspace{0.3em}
\noindent
\textbf{Reducing Computational Demands.}
The computational intensity of training and deploying NeRF models, especially for high-resolution medical imaging data, poses a significant challenge \cite{chen2024nerfhub}. Future research should aim to optimize these models for greater efficiency, potentially through algorithmic improvements, the use of more efficient neural network architectures, or leveraging hardware accelerations \cite{zhang2024federated}.


\noindent
\textbf{Integration with Diagnosis.}
Exploring the integration of NeRF-generated 3D models into diagnostic workflows, surgical planning, and even real-time image guidance during procedures could have profound implications for patient care. This could involve developing tools and interfaces that allow clinicians to interact with and manipulate 3D reconstructions, improving treatment planning and execution.

\noindent
\textbf{Enhancing Real-Time Performance.}
NeRFs in medical imaging are poised for significant advancements, particularly with the integration of new techniques. NeRFs could integrate with 3D Gaussian Splatting \cite{kerbl20233d} for real-time radiance field rendering, which promises to address some of the current limitations of NeRFs, such as the high computational cost and the need for faster rendering times without compromising the quality of the 3D reconstructions \cite{cai2024radiative,nikolakakis2024gaspct}. 
The computational intensity of NeRFs currently limits their application in time-sensitive scenarios such as emergency medicine or real-time surgical assistance. 
By incorporating 3D Gaussian Splatting, the computational load could be significantly reduced, making NeRF-based applications more accessible, including mobile devices used in remote diagnostics.

\noindent
\textbf{Agentic NeRFs.} 
The integration of NeRFs with Reinforcement Learning (RL) \cite{wang2023deep} in medical imaging represents a forward-looking approach that can significantly improve the efficiency, accuracy, and personalization of medical imaging and related procedures. 
By learning from interactions between data and environments, RL models can optimize NeRFs to meet specific needs of medical imaging, addressing challenges such as reducing scan time, improving patient comfort, and simplifying the interpretation of medical images \cite{driess2022reinforcement}. 

%

\noindent
\textbf{Foundation Models of NeRFs.}  Foundation Models (FMs) \cite{bommasani2021opportunities} offer a promising approach to address some of the unique challenges faced in medical image analysis and reconstruction. 
The integration of NeRFs with FMs in medical imaging could lead to significant advancements in accuracy, efficiency, and generalizability across different medical imaging modalities and clinical applications. 
FMs are trained on vast datasets, enabling them to learn a wide range of features and patterns \cite{zhou2023foundation}. 
When applied to medical imaging, NeRFs can help overcome the challenge of data variability across different medical conditions and imaging modalities. 
The depth and complexity of FMs allow them to capture high-level abstractions and intricate details from medical images \cite{zhang2023challenges}. 
When used to generate or refine NeRFs, these models can potentially improve the accuracy of 3D reconstructions, enhance the detection of diseases, and provide more precise anatomical models for simulation purposes.



\section{Conclusion}
\label{conclusion}

In conclusion, the exploration of Neural Radiance Fields (NeRF) in medical imaging presents a promising pathway for enhancing diagnostic precision, treatment planning, and early disease detection. 
This paper has examined the multifaceted challenges of applying NeRFs to medical image data, such as adapting to the distinct imaging principles of medical modalities, capturing intricate internal structures essential for accurate diagnostics, defining object boundaries with precision, and interpreting color and density variations critical for medical analysis. 
Through a comprehensive examination of current methodologies and an discussion of the limitations and potential of NeRFs, we emphasize the necessity for innovative solutions that can navigate these obstacles. 






\small
\bibliographystyle{IEEEbib}
\bibliography{refs}

\end{document}